\documentstyle[aps,12pt]{revtex}
\begin{document}
\title{Almost non-magnetic ground state of the V$^{4+}$ ion:\\
the case of BaVS$_3$}
\author{Z.~Ropka$^{a}$ and R.J. Radwanski$^{a,b}$}
\address{$^a$Center for Solid State Physics, S$^{nt}$ Filip 5, 31-150 Krak%
\'{o}w.\\
$^{b}$Institute of Physics, Pedagogical University, 30-084 Krak\'{o}w,\\
POLAND.}
\maketitle

\begin{abstract}
We have shown that the V$^{4+}$ ion with 1 d electron can have almost
non-magnetic ground state under the action of the octahedral crystal field
in the presence of the spin-orbit coupling and off-cubic distortions. Such
the situation occurs in the hexagonal BaVS$_3$. The fine discrete electronic
structure with the weakly-magnetic Kramers-doublet ground state is the
reason for anomalous properties of BaVS$_3$.


\end{abstract}

The V$^{4+}$ ion is realized in such compounds as CaV$_{4}$O$_{9},$ MgVO$%
_{3} $, (VO)$_{2}$P$_{2}$O$_{7}$, BaVS$_{3}$ that become nowadays very
popular [1-9]. The V$^{4+}$ ion with one 3d electron, a 3d$^{1}$ system, is
usually treated as S=1/2 system i.e. with the spin-only magnetism and with
taking into account the spin degrees of freedom only [1-9]. The neglect in
the current literature of the orbital moment is consistent with the
widely-spread conviction that the orbital magnetism plays rather negligible
role due to the quenching of the orbital moment for 3d ions. The above
mentioned compounds are example of numerous compounds in which this S=1/2
behavior is drastically violated [1-9]. One of this drastic violation
experimentally-observed is associated with the substantial departure of the
temperature dependence of the paramagnetic susceptibility from the Curie law
at low temperatures and the formation of a non-magnetic state at low
temperatures.

The aim of this Letter is to study the magnetic moment of the V$^{4+}$ ion
in BaVS$_3$ by consideration of the eigen-states and eigen-functions of the
3d$^1$ electron system under the action of the octahedral crystal field
(CEF)\ in the presence of the spin-orbit (s-o) coupling and a small
off-cubic distortion. It turns out that, despite of the Kramers doublet
ground state, a ground state with quite small magnetic moment can be
obtained as an effect of the off-cubic distortions.

One 3d electron is described by L=2 and S=1/2. The term $^2$D is 10-fold
degenerated. Its degeneracy in a solid is removed by crystal-field
interactions and the intra-atomic spin-orbit interactions. The latter is
described by a well-known Hamiltonian H$_{s-o}$ = $\lambda L\cdot S$. The
crystal-field Hamiltonian is related with the local symmetry of the V$^{4+}$%
-ion site. This local symmetry can be quite low as well as there can be a
few cationic sites. In case of the hexagonal structure of BaVS$_3$ (the
CsCoCl$_3$ type) the nearest cationic surrounding is formed by 6 sulphur
anions. This sulphur surrounding is close to the octahedral surrounding
[7,8] - the hexagonal axis of the elementary cell lies along the main
diagonal of the local octahedron. Thus, the crystallographic structure of
BaVS$_3$ suggests the start for the analysis of the crystal-field
interactions from the dominant octahedral CEF\ interactions supplemented by
an off-cubic trigonal distortion.

This situation can be exactly traced by the consideration of a
single-ion-like Hamiltonian

\begin{center}
$H_d=H_{CF}^{octa}+H_{s-o}+H_{CF}^{tr}=B_4(O_4^0+5O_4^4)+\lambda L\cdot
S+B_2^0O_2^0+\mu _B(L+$g$_eS)\cdot B_{ext}(1)$
\end{center}

in the 10-fold degenerated spin+orbital space. The last term allows to
calculate the influence of the external magnetic field. It allows also the
calculations of the paramagnetic susceptibility. Such type of the
Hamiltonian has been widely used in analysis of electron-paramagnetic
resonance (EPR)\ spectra of 3d-ion doped systems [10,11]. Here we use this
Hamiltonian for systems, where the 3d-ion is the real part of the crystal.

The resulting electronic structure of the V$^{4+}$ ion contains 5 Kramers
doublets separated in case of the dominant cubic CEF interactions into 3
lower doublets (originated from the T$_{2g}$ cubic sub-term) and 2 doublets
(from the E$_{g}$ sub-term) 2-3 eV above (T$_{2g}$-E$_{g}$ = 120$\cdot $B$%
_{4}$; we take B$_{4}$= +200 K, the sign + corresponds to the ligand
octahedron). The 3 lower doublets are spread over 1.5$\cdot \lambda _{s-o}$ (%
$\lambda _{s-o}$ is taken as +360 K, after [10]). The octahedral CEF\ states
in the presence of the spin-orbit coupling have been shown in Ref. 12 - for
the 3d$^{1}$ system the T$_{2g}$ subterm is split into lower quartet and a
doublet at 1.5$\cdot \lambda _{s-o}$. The quartet states, that are in fact
two Kramers states, have very small magnetic moments [12]. Negative values
of the trigonal distortion parameter B$_{2}^{0}$ yields the ground state
that is almost non-magnetic. Here we focus on the trigonal distortion as the
tetragonal distortion has been studied in Ref. 13. For B$_{2}^{0}$ = -20 K
the ground state moment amounts to $\pm 0.06$ $\mu _{B}$. It is composed
from the orbital moment of $\pm 0.74$ and the spin moment of $\mp 68$ $\mu
_{B}$ (antiparallel). The sign $\pm $ corresponds to 2 Kramers conjugate
states. The excited Kramers doublet lies at 110 K (=10 meV) and is strongly
magnetic - its moment amounts to $\pm $0.45 $\mu _{B}$ (=$\pm $0.46+2$\cdot $%
($\mp $0.005)). We ask ourselves for the origin of the trigonal distortion.
Obviously, it has the main source in the lattice trigonal distortion. In
fact, it can be quite easily realized in the hexagonal structure via the
elongation or the contraction (stretching) along the hexagonal axis without
the change of the overall hexagonal symmetry. The trigonal distortion of the
crystal field, with the negative value of B$_{2}^{0}$, is additionally
produced by the positive charges located on the V ions in the hexagonal
plane.

The existence of such the discrete electronic structure with such strange
magnetic characteristics causes enormous effect in the temperature
dependence of magnetic and electronic properties like it was calculated for
LaCoO$_{3}$ [14]. In particular, the susceptibility is very small at low
temperatures and exhibits a maximum at ambient temperatures. Moreover, we
would like to point out that the Kramers spin-like degeneracy of the ground
state has to be removed somewhere - we expect it to occur at very low
temperatures. Thus our theory predicts BaVS$_{3}$ to exhibit heavy-fermion
like properties in the specific heat at ultra-low temperatures.

In conclusion, the very strong influence of the spin-orbit coupling on the
realized ground state and its magnetic moment has been proved for the V$%
^{4+} $ ion within the CEF theory. Almost non-magnetic Kramers doublet
ground state can be formed by off-cubic crystal-field distortions. Such
conditions are realized in the hexagonal structure of BaVS$_{3}$. The
crystal-field effect of a low symmetry and the spin-orbit coupling is,
according to us, the reason for experimentally-observed anomalous behavior
of BaVS$_{3}$. In our atomic-like approach there is very strong correlation
between the local symmetry and the realized magnetic moment. This
atomic-like approach, extended to the quantum atomistic solid-state theory
(QUASST), points out that such considerations are physically meaningful as
the whole crystal BaVS$_{3}$ is built up from the face-sharing VS$_{6}$
octahedra stacking along the hexagonal \ axis. In order to avoid a
undeserved critique we would like to say that we do not claim that
everything can be explained in a solid state by atomic physics but more
subtle effects known from atomic physics have to be employed for the
understanding of solid-state properties. It means, in particular, that
compounds with V$^{4+}$ ions cannot be simplified as S=1/2 systems, despite
of consideration of spin ladders or spin chains. It means also that the
intra-atomic spin-orbit coupling has to be taken into account for any
meaningful description of electronic and magnetic properties of 3d-ion
containing compounds.

\end{document}